\documentclass{article}
\usepackage{hiph-art}
\usepackage{graphicx}
\newcommand{\J}{J/\psi}
\newcommand{\NJ}{N_{\J}}

\newcommand{\Ncoll}{N_{coll}}

\newcommand{\ccbar}{c \bar c}
\newcommand{\Nccbar}{N_{\ccbar}}

\newcommand{\pts}{\langle{p_T}^2\rangle}

\newcommand{\pt}{ $p_T$ }
\volnumber{XX} \issuenumber{Y} \edyear{2006}                             
\frompage{1} \topage{8}                                                  
\recrevdate{November 24, 2005}                                           
\def\rightmark{Charm Quark Thermalization}
\def\leftmark{R. Thews}
 
\title{Formation of charmonium states in heavy ion collisions and thermalization
of charm} 
 
\authors{ 
{R. L. Thews
\index{Thews, R.} 
}\\[2.812mm]
{\normalsize
Department of Physics, University of Arizona,\\ 
Tucson, AZ 85721, USA\\[0.2ex] 
}}
 
\abstract{
We examine the possibility to utilize in-medium charmonium formation in heavy
ion interactions at collider energy as a probe of the properties of 
the medium.  This is possible because the formation process involves
recombination of charm quarks which imprints a signal on the resulting
normalized transverse momentum distribution containing information
about the momentum distribution of the quarks.
We have contrasted the
 transverse momentum spectra of $\J$, characterized by $\pts$, which result
from the formation process in which the
charm quark distributions are taken at opposite limits with regard to
thermalization in the medium.  The first uses charm quark distributions 
unchanged from their initial production
in a pQCD process, appropriate if their interaction with the medium is
negligible.  The second uses charm quark distributions which are in complete
thermal equilibrium with the transversely expanding medium, appropriate 
if a very strong interaction between charm quarks and medium exists.  
We find that the resulting $\pts$ of the formed $\J$ should allow one to
differentiate between these extremes, and that this differentiation is
not sensitive to variations in the detailed dynamics of in-medium formation.
We include a comparison of predictions of this model with preliminary PHENIX 
measurements, which indicates
compatibility with a substantial fraction of in-medium formation.}  

 
\makeindex
\begin{document}
 
\maketitle


\section{Introduction}\label{intro}

The in-medium formation picture we consider here \cite{Thews:2000rj}
uses competing formation
and dissociation reactions in a Boltzmann equation to calculate the final $\J$ 
population.  The absolute value of this formation was found to be 
very sensitive to the underlying
charm quark momentum distributions \cite{Thews:2001hy}.  There
is also quite strong variation of the results which depend on 
largely unconstrained model parameters involving details of  
the size and expansion profile of the deconfinement
region.   
The initial PHENIX data \cite{Adler:2003rc}
suffered from low statistics, and was compatible with 
a fairly large region of model parameter space \cite{Thews:2003da}.

Recent work in this area concentrated on finding a signature 
for in-medium $\J$ formation which is
independent of the detailed dynamics and magnitude of
the formation.  We found that the \pt spectrum of the 
formed $\J$ may provide such a signature \cite{Thews:2005vj}.

The calculations which used  
initial charm quark 
momentum distributions from NLO pQCD amplitudes to generate a sample
of $\ccbar$ pairs, were then supplemented
by an initial-state transverse momentum kick to simulate confinement and 
nuclear effects.  
In the evolution of the interacting
system size from pp to pA to AA collisions, the \pt will be
in general increased due to initial-state effects of interaction
of constituents in the nuclei. 
\begin{equation}
\pts_{AB} - \pts_{pp}\; = \lambda^2 \;[\bar{n}_A + \bar{n}_B- 2], 
\label{AApt}
\end{equation}
where $\bar{n}_A$ is the impact-averaged number of inelastic interactions
of the projectile nucleons in nucleus A, and $\lambda^2$ is the square of the
transverse momentum transfer per collision. 
The PHENIX measurements of $\J$ \pt spectra in pp and minimum-bias 
d-Au interactions \cite{Adler:2005ph}
allow us to determine the amount of initial state $k_T$ needed to
supplement our collinear pQCD events.
(This is equivalent to using a hadronization model similar to a
color evaporation scheme, except that we assume that the resulting
\pt of the resulting $\J$ is determined by the pair \pt for all invariant
mass of the combinations. One can then extrapolate to
Au-Au and predict the spectrum of $\J$ which are produced from 
hadronization of the initial "diagonal" $\ccbar$ pairs,
again for minimum bias interactions.  (We use diagonal
to distinguish these pairs from the "off-diagonal" combinations which
contribute to in-medium $\J$ formation.)
One finds
$\bar{n}_A = 5.4$ from minimum bias d-Au interactions at RHIC
energy (using $\sigma_{pp}$ = 42 mb), which leads to
$\lambda^2 = 0.35 \pm 0.14\ GeV^2$.  We note that the relatively large
uncertainty comes entirely from the difference in $p_T$ broadening
in d-Au between positive and negative rapidity.

\begin{figure}[h]
  \includegraphics[clip, height=.45\textheight]{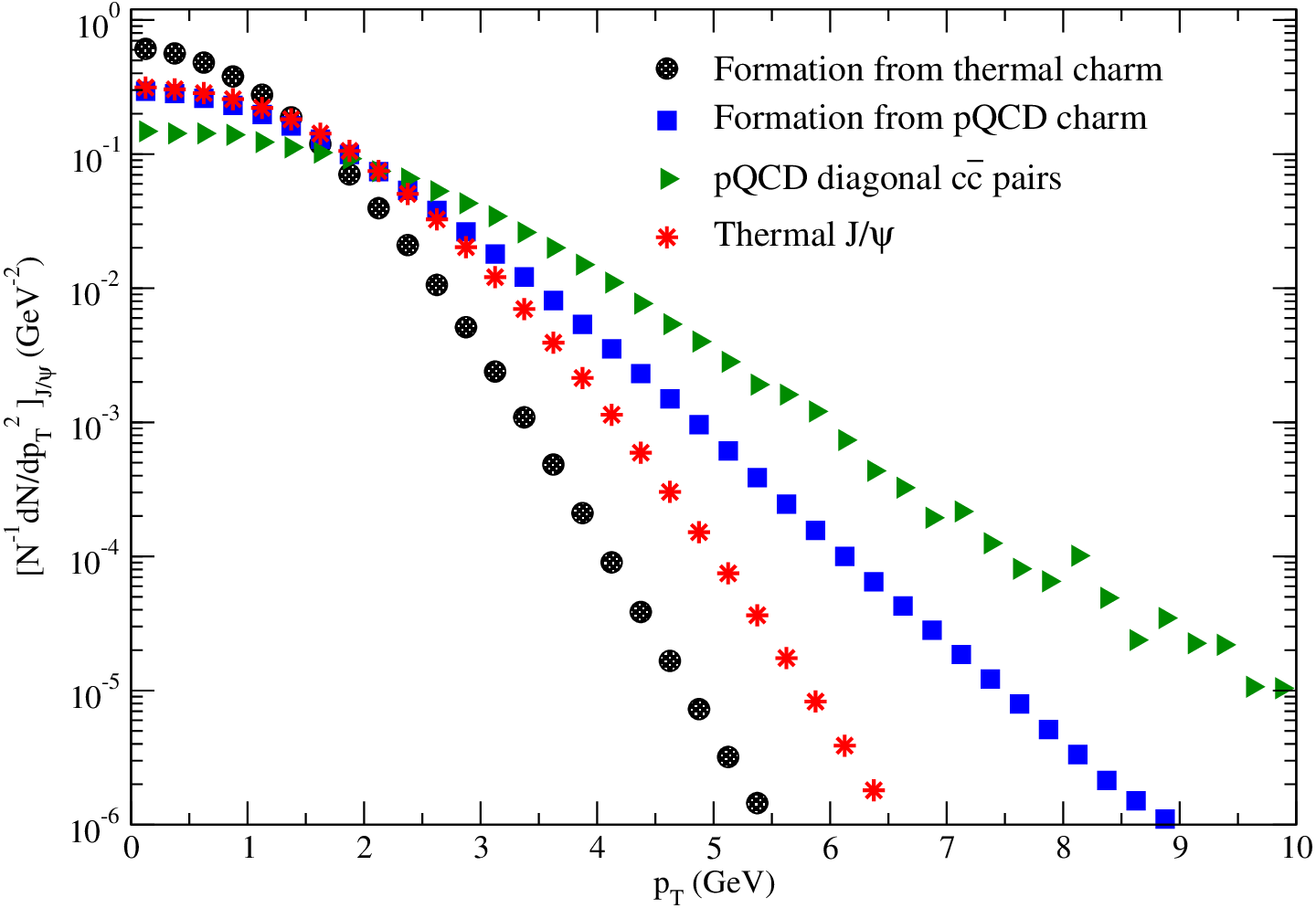}
  \caption{Comparison of in-medium $\J$ transverse momentum spectra
           predictions from various scenarios.}
\label{jpsiptallpredictions}
\end{figure}

Our prediction for the "normal" evolution of the \pt spectrum in Au-Au
interactions is shown by the triangular points in Fig.\ref{jpsiptallpredictions}.

The properties of the normalized pt spectrum for the formation process  
follow from two separate effects:  First, the fact that the process 
is dominated by
the off-diagonal pairs introduces a modified initial \pt distribution.
Next, one weights these pairs by a formation probability for $\J$.  We use
the operator-product motivated cross section  
for $\ccbar$ forming $\J$
with emission of a final-state gluon, which of course is just the
inverse of the dissociation process.  However, any cross section which
has the same general properties as this one gives essentially the
same result \cite{Thews:2005vj}.
We show by the square points in 
Fig.\ref{jpsiptallpredictions} the prediction for the formed $\J$.
 One sees that this spectrum
 is substantially narrower
than the one with no in-medium formation. 

The same procedure was employed when using quark momentum distributions
which follow for charm in thermal equilibrium with the expanding
region of deconfinement.
The parameters of temperature and 
maximum transverse expansion rapidity are determined by a fit to
this thermal behavior of the produced light hadrons. The application 
to charm quarks was originally motivated in Ref. \cite{Batsouli:2002qf}, 
who showed that
the low-\pt spectrum of decay leptons from charmed hadron decays would not
be able to differentiate between the thermal and a purely pQCD
distribution.   We show here, however, that the \pt spectrum of
in-medium formed $\J$ is very sensitive to this distribution.  
The
circles in Fig.\ref{jpsiptallpredictions} result from formation 
calculations using T = 170 MeV and
$y_{Tmax}$ = 0.5 for the thermal charm quarks.
One sees that this \pt spectrum
is narrower yet than in-medium formation from pQCD quarks. 
Finally, we show by the stars 
the \pt spectrum of $\J$ which themselves obey this thermal
distribution.  The resulting spectrum falls between the in-medium 
formation spectra for either pQCD or thermal charm quark momentum 
distributions.   

\section{Centrality behavior}

\begin{figure}[h]
  \includegraphics[clip, height=.45\textheight]{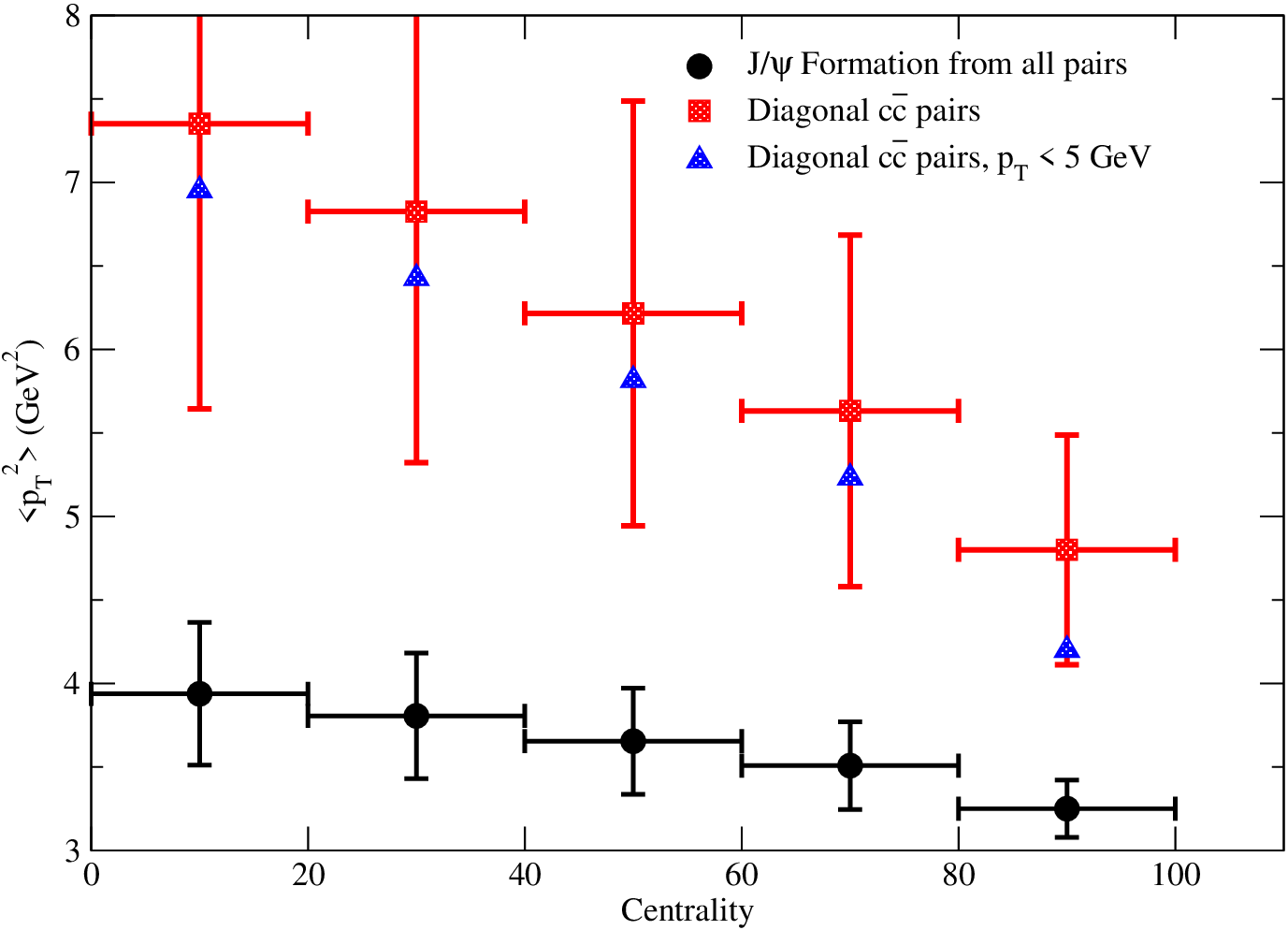}
  \caption{Centrality dependence of $\J \pts$ contrasting predictions 
           assuming either 100 \% production from initial $\ccbar$ 
           pairs or 100 \% in-medium formation.}
\label{jpsiptwidthsvscentrality}
\end{figure}

We now proceed to investigate the variation of the pQCD-based results with
respect to the collision centrality in Au-Au interactions.  First, we
use the value of $\lambda^2$ extracted from pp and pA data, together with
values of $\bar{n}_A$ calculated as a function of collision centrality, to 
recalculate the $\pts$ values for either the initial production or the
in-medium formation separately.  This provides the centrality behavior of
the $\J$ spectrum in the case that one or the other of these mechanisms is
solely responsible for the total $\J$ population.  We show these results together 
in Fig. \ref{jpsiptwidthsvscentrality}. 
One sees as expected that $\pts$ is maximum for the most central 
collisions, but the absolute magnitudes are widely separated for
initial production and in-medium formation at each centrality.  One should note
that the uncertainties are dominated by the difference between the 
\pt-broadening measurements at positive and negative rapidities
in the d-Au interactions.  Thus the point-to-point uncertainties
are much smaller for the centrality behavior.  We have also
included separate values for $\pts$ in the region limited by
a maximum \pt  of 5 GeV, to facilitate comparison with experiment in this same range.

\begin{figure}[h]
  \includegraphics[clip, height=.45\textheight]{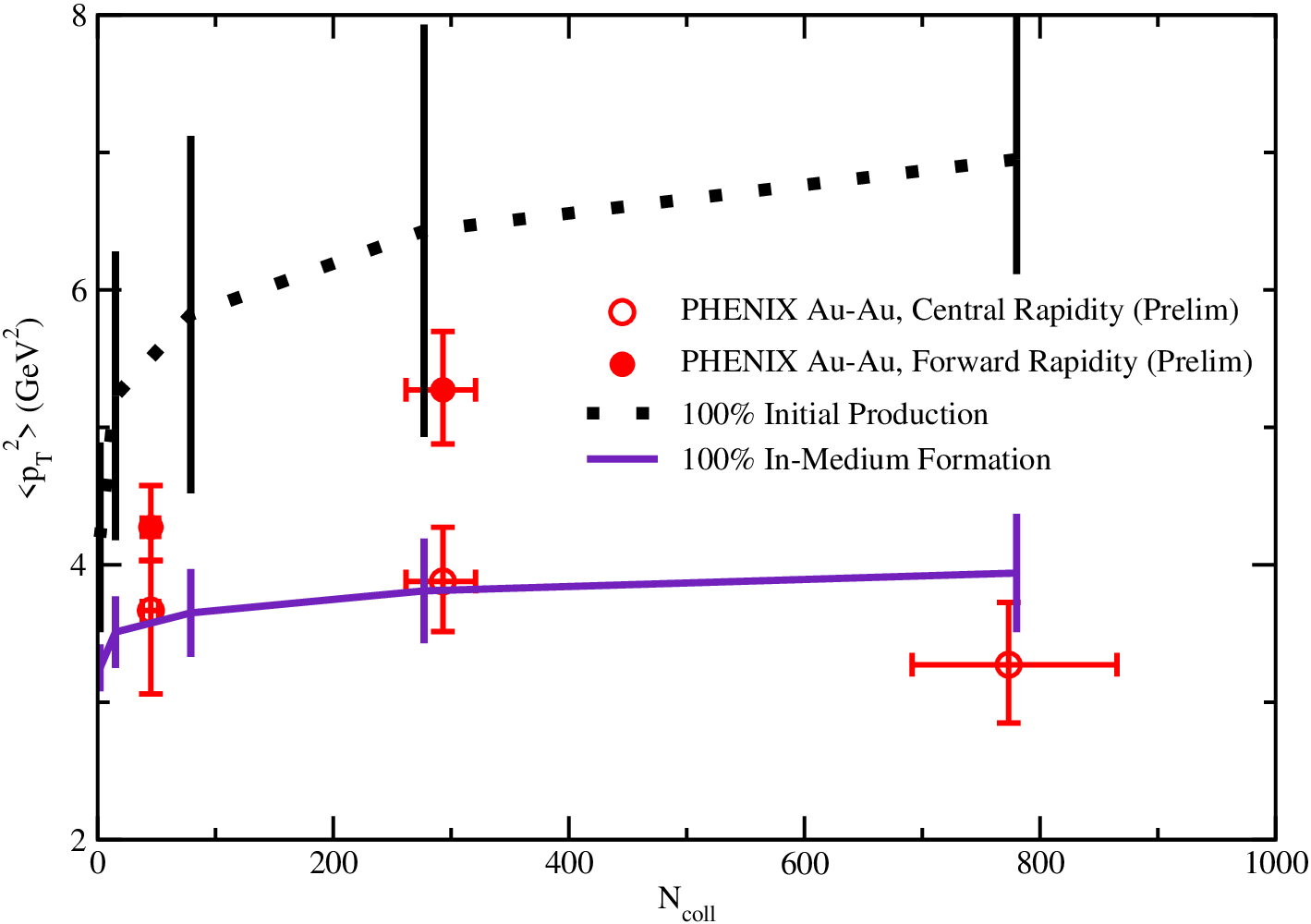}
  \caption{Comparison of $\pts$ 
        predictions for initially-produced $\J$ and in-medium
           formation with preliminary PHENIX
           measurements.  The collision centrality is parameterized by
           the number of nucleon-nucleon collisions $\Ncoll$.  The theory 
          curves and uncertainties are taken from Fig. \ref {jpsiptwidthsvscentrality}.}
\label{pt2vsncollanddata}
\end{figure}

There exist preliminary results from PHENIX for $\pts$ of $\J$ produced
in Au-Au collisions \cite{PereiraDaCosta:2005xz}.  These are reported as a function
of the number of nucleon-nucleon collisions $\Ncoll$.  In order to compare
with our predictions, we transform centrality to $\Ncoll$ using a
Glauber model.  The resulting predictions are shown in 
Fig. \ref{pt2vsncollanddata}.  
Although there is substantial uncertainty in the absolute values, one
can infer a clear preference for in-medium formation over the initial
production prediction. 

\begin{figure}[h]
  \includegraphics[clip, height=.45\textheight]{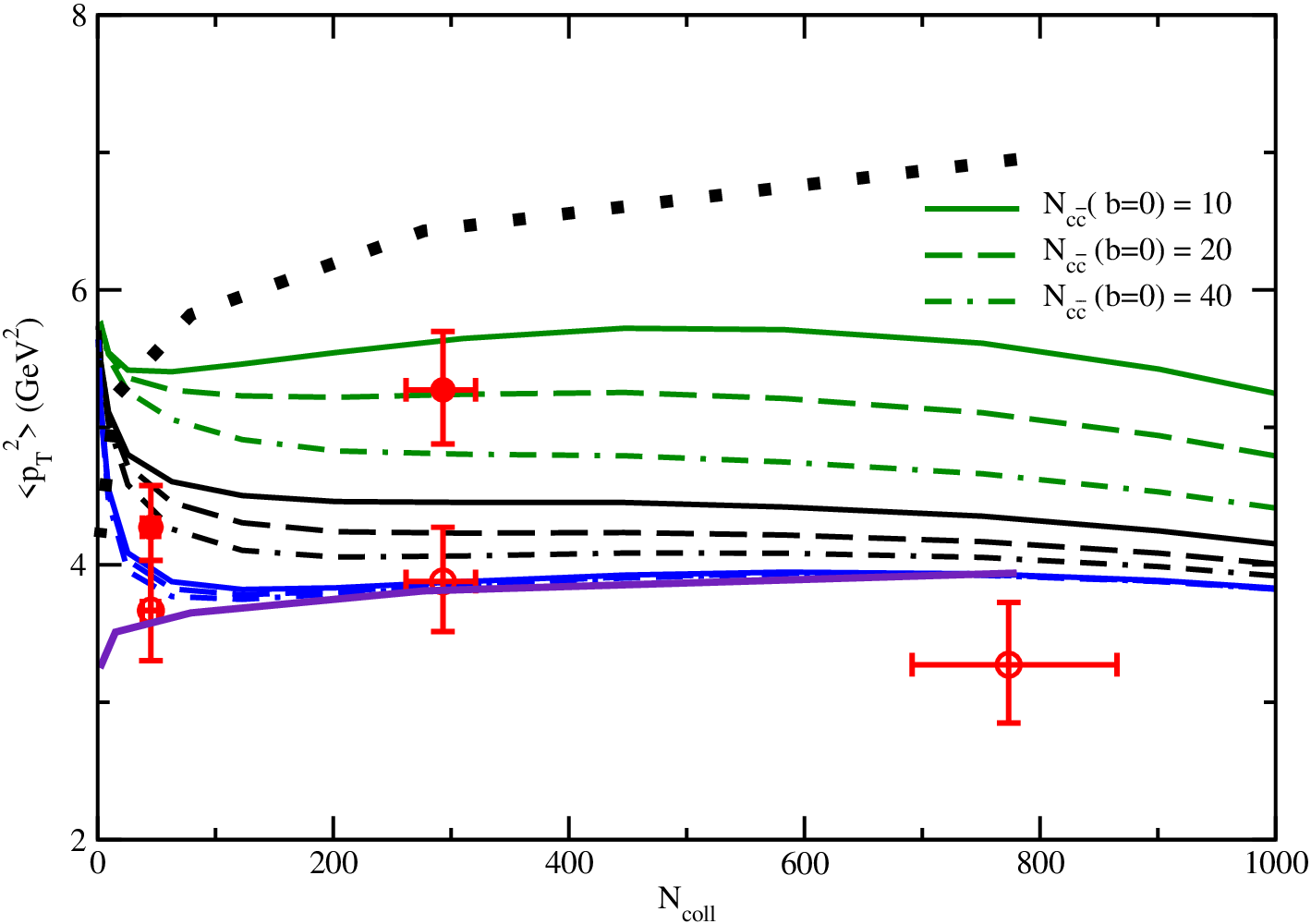}
  \caption{Variation of predictions in Fig. \ref{pt2vsncollanddata} resulting
           from a model calculation of the relative fraction of $\J$ 
           which originate from initial production or in-medium formation.}
\label{pt2vsncollinitialandform}
\end{figure}

In order to provide a meaningful prediction for the overall
 $\J$
spectrum, one should of course include both initial production and
in-medium formation together as sources.  This requires some estimate of the
relative magnitudes of these processes, and is subject to considerable
model uncertainties.  What we can say, however, is that in-medium formation
will be most dominant for central collisions, where the quadratic dependence
on $\Nccbar$ is enhanced.  Conversely, one expects that initial production
will increase in relative importance for very peripheral collisions.  To get
an approximate idea of how this effect will appear, we revert to our
original model calculations which included the absolute magnitude results 
\cite{Thews:2001hy}.  One relevant parameter is the number of initial
pairs, $\Nccbar$, parameterized by its value at zero impact parameter.
These results are
shown in Fig. \ref{pt2vsncollinitialandform} for three representative
values of $\Nccbar$ = 10, 20, and 40, which span the range of values extracted
from STAR \cite{Adams:2004fc} and 
PHENIX \cite{Adler:2004ta} measurements of charm inferred from
semileptonic decays and direct reconstruction of D mesons.

The 100 \% curves for initial production and in-medium formation remain
as in Fig. \ref{pt2vsncollanddata}.  Between these two extremes are shown
the calculations including centrality-dependent fractions of each, as
described above.  Each $\Nccbar$ value is shown by the solid, dashed, 
and dot-dashed lines, respectively.  The duplication of these lines 
results from different initial temperature values of 300, 400, and 500 MeV,
which provides variation in the gluonic dissociation rates and the
deconfinement lifetime in the model calculations.  One sees that, with
the exception of extremely peripheral collisions, all of the predictions
remain substantially below that for initial formation alone.  It would
require a substantial reduction of experimental uncertainty in the 
measured $\pts$, however, to pin down the preferred model parameters
appropriate for in-medium formation.

\section{Hadronization Contribution}

\begin{figure}[h]
  \includegraphics[clip, height=.45\textheight]{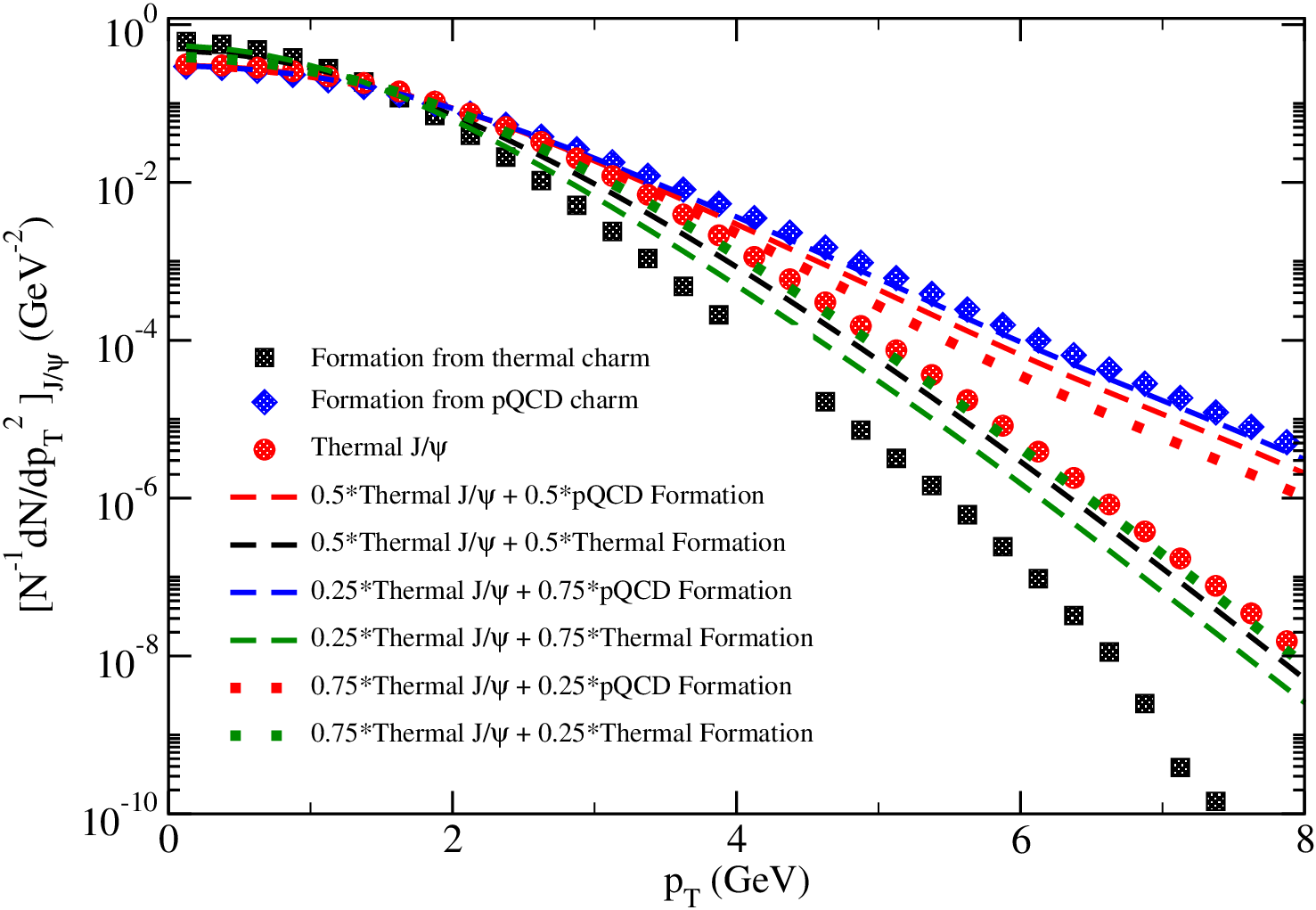}
  \caption{Comparison of $p_T$ spectra for $\J$ resulting from variable
           fractions of in-medium formation and subsequent 
           hadronization of residual charm.}
\label{formandhadronizationcombination}
\end{figure}

In addition to in-medium formation of $\J$, one needs to consider 
the possibility of subsequent production at the hadronization transition.
This contribution must be significant, as can be confirmed by an examination
of the absolute values of in-medium formation in the model calculations.
For central collisions, our calculated values of the ratio $\NJ / \Nccbar$ 
typically range from 0.004 to 0.01 as the model parameters are varied.
This means that essentially all of the initial $\ccbar$ pairs will survive
through the deconfined phase and some fraction of these must hadronize into
additional $\J$.  For an estimate of this fraction, we utilize the
statistical hadronization model \cite{Braun-Munzinger:2000px}. 
The most recent applications
of this model to RHIC Au-Au collisions 
\cite{Andronic:2003zv,Bratkovskaya:2004cq} predict
values for $\NJ / \Nccbar
\approx 0.005$, comparable to that for our in-medium formation.  Thus
we need to consider contributions to the final $\J$ population which
originate in approximately equal amounts from each mechanism.  Fortunately
one can make this combination for central collisions alone, since 
both in-medium formation and statistical hadronization predict
the same centrality behavior \cite{Thews:2001hy}.
The results of such a calculation are shown in Fig. 
\ref{formandhadronizationcombination}.
 
Shown are the $p_T$ spectra for the in-medium formation utilizing 
pQCD charm distributions (diamonds) and thermal plus flow charm
distributions (squares).  They are at the extremes of $\pts$, as previously
noted.  Also shown is the thermal plus flow distribution for a
direct $\J$, which we take as our expectation if statistical hadronization
is the dominant source of $\J$.  The distributions which fall between
the extremes are the result of various (but comparable) contributions
of in-medium formation and hadronization.  One sees that these combinations
predict the spectra evolve toward that for the thermal $\J$ as its fraction
increases.  However, the rate of this evolution is substantially greater
for in-medium formation from thermal charm quarks than the corresponding
behavior for in-medium formation from pQCD charm quarks.  Thus the possibility
to utilize the $p_T$ spectra as a probe of charm thermalization in
the medium remains a viable option.



\section*{Acknowledgments}
This research was partially supported by the U.S. Department of
Energy under Grant No. DE-FG02-04ER41318.


\vfill\eject

\begin{thebibliography}{99}



\bibitem{Thews:2000rj}
R.~L.~Thews, M.~Schroedter and J.~Rafelski,
Phys.\ Rev.\ C {\bf 63}, 054905 (2001)
[arXiv:hep-ph/0007323].


\bibitem{Thews:2001hy}
R.~L.~Thews,
Nucl.\ Phys.\ A {\bf 702}, 341 (2002)
[arXiv:hep-ph/0111015].


\bibitem{Adler:2003rc}
S.~S.~Adler {\it et al.}  [PHENIX Collaboration],
Phys.\ Rev.\ C {\bf 69}, 014901 (2004)
[arXiv:nucl-ex/0305030].

\bibitem{Thews:2003da}
R.~L.~Thews,
J.\ Phys.\ G {\bf 30}, S369 (2004)
[arXiv:hep-ph/0305316].


\bibitem{Thews:2005vj}
  R.~L.~Thews and M.~L.~Mangano,
  Phys.\ Rev.\ C {\bf 73}, 014904 (2006)
  [arXiv:nucl-th/0505055].


\bibitem{Adler:2005ph}
  S.~S.~Adler {\it et al.}  [PHENIX Collaboration],
  Phys.\ Rev.\ Lett.\  {\bf 96}, 012304 (2006)
  [arXiv:nucl-ex/0507032].


\bibitem{Batsouli:2002qf}
  S.~Batsouli, S.~Kelly, M.~Gyulassy and J.~L.~Nagle,
  Phys.\ Lett.\ B {\bf 557}, 26 (2003)
  [arXiv:nucl-th/0212068].

\bibitem{PereiraDaCosta:2005xz}
  H.~Pereira Da Costa  [PHENIX Collaboration],
  arXiv:nucl-ex/0510051.

\bibitem{Adams:2004fc}
J.~Adams {\it et al.}  [STAR Collaboration],
arXiv:nucl-ex/0407006.

\bibitem{Adler:2004ta}
S.~S.~Adler {\it et al.}  [PHENIX Collaboration],
arXiv:nucl-ex/0409028.


\bibitem{Braun-Munzinger:2000px}
P.~Braun-Munzinger and J.~Stachel,
Phys.\ Lett.\ B {\bf 490}, 196 (2000)
[arXiv:nucl-th/0007059].

\bibitem{Andronic:2003zv}
  A.~Andronic, P.~Braun-Munzinger, K.~Redlich and J.~Stachel,
  Phys.\ Lett.\ B {\bf 571}, 36 (2003)
  [arXiv:nucl-th/0303036].

\bibitem{Bratkovskaya:2004cq}
  E.~L.~Bratkovskaya, A.~P.~Kostyuk, W.~Cassing and H.~Stoecker,
  Phys.\ Rev.\ C {\bf 69}, 054903 (2004)
  [arXiv:nucl-th/0402042].

\end{thebibliography}
\end{document}